\documentclass{article}

\usepackage{times}
\usepackage{amssymb,amsmath,amsthm}
\usepackage{mathrsfs} 
\usepackage{epsfig}
\usepackage{color}

\usepackage{xy}

\newcommand{\ie}{\emph{i.e.}~}
 
\newcommand{\cf}{\emph{cf.}~}

\def\A{\mathcal A}
\def\B{\mathscr B}
\def\BB{\mathcal B}
\def\C{\mathbb C}
\def\D{\mathcal D}

\def\F{\mathscr F}

\def\G{\mathcal G}
\def\H{\mathcal H}
\def\K{\mathcal K}

\def\M{\mathsf M}
\def\N{\mathbb N}

\def\R{\mathbb R}

\def\U{\mathcal U}

\def\Z{\mathbb Z}

\def\<{\left\langle} 
\def\>{\right\rangle} 
\def\({\left(} 
\def\){\right)} 
\def\[{\left[} 
\def\]{\right]} 
\def\ltwo{\mathsf{L}^{\:\!\!2}}
\def\lone{\mathsf{L}^{\:\!\!1}}
\def\lp{\mathsf{L}^{\:\!\!p}} 

\def\e{\mathop{\mathrm{e}}\nolimits} 
\def\d{\mathrm{d}}
\def\Aut{\mathop{\mathrm{Aut}}\nolimits}
\def\supp{\mathop{\mathrm{supp}}\nolimits}

\def\diag{\mathop{\mathrm{diag}}\nolimits}

\def\Hom{\mathrm{Hom}}

\def\SU{\textsc{SU}}

\newtheorem{Theorem}{Theorem}[section] 
\newtheorem{Remark}[Theorem]{Remark} 
\newtheorem{Lemma}[Theorem]{Lemma}
\newtheorem{Corollary}[Theorem]{Corollary} 
\newtheorem{Proposition}[Theorem]{Proposition} 
\newtheorem{Definition}[Theorem]{Definition} 
\newtheorem{Example}[Theorem]{Example}
\newtheorem{Procedure}[Theorem]{Procedure}

\begin{document}

%-------------------------------------------------------------------------------------------------------------------
% Title 
%-------------------------------------------------------------------------------------------------------------------

\title{\Large\textbf{Spectral analysis for convolution operators on locally compact groups}}
   
\author{M. M\u antoiu$^1$ and R. Tiedra de Aldecoa$^2$} 
\date{\small}
\maketitle
\vspace{-1cm}

\begin{quote}
\emph{
\begin{itemize}
\item[$^1$] Institute of Mathematics ``Simion Stoilow'' of the Romanian Academy,\\
P.\,O. Box 1-764, 014700 Bucharest, Romania
\item[$^2$] D\'epartement de math\'ematiques, Universit\'e de Cergy-Pontoise,\\
2, Avenue Adolphe Chauvin, 95302 Cergy-Pontoise Cedex, France 
\item[] \emph{E-mails:} Marius.Mantoiu@imar.ro, rafael.tiedra@u-cergy.fr
\end{itemize}
  }
\end{quote}

%-------------------------------------------------------------------------------------------------------------------
% Abstract 
%-------------------------------------------------------------------------------------------------------------------

\begin{abstract} 
We consider operators $H_\mu$ of convolution with measures $\mu$ on locally compact groups. We characterize the
spectrum of $H_\mu$ by constructing auxiliary operators whose kernel contain the pure point and singular subspaces
of $H_\mu$, respectively. The proofs rely on commutator methods.
\end{abstract}



%-------------------------------------------------------------------------------------------------------------------
\section{Introduction}
\setcounter{equation}{0}
%-------------------------------------------------------------------------------------------------------------------

Any selfadjoint operator $H$ in a Hilbert space $\H$, with spectral measure $E_H$ and spectrum $\sigma(H)$, is
reduced by an orthogonal decomposition 
$$
\H=\H_{\rm ac}(H)\oplus\H_{\rm p}(H)\oplus\H_{\rm sc}(H),
$$
that we briefly recall (\cf \cite[Sec. 7.4]{Weidmann80}). Denote by ${\rm Bor}(\R)$ the family of Borel subsets of
$\R$. Then, for any $f\in\H$, one has the positive Borel measure 
$$
\nu^f_H:{\rm Bor}(\R)\rightarrow[0,\infty),\quad A\mapsto\nu^f_H(A):=\left\|E_H(A)f\right\|^2=\<f,E_H(A)f\>.
$$
We say that $f$ \emph{belongs to the spectral subspace $\H_{\rm p}(H)$} if $\nu^f_H$ is pure point, $f$
\emph{belongs to the spectral subspace $\H_{\rm ac}(H)$} if $\nu^f_H$ is absolutely continuous with respect to
the Lebesgue measure, and $f$ \emph{belongs to the spectral subspace $\H_{\rm sc}(H)$} if $\nu^f_H$ is singularly
continuous with respect to the Lebesgue measure. One also uses the notations
$\H_{\rm c}(H):=\H_{\rm ac}(H)\oplus\H_{\rm sc}(H)$ for the \emph{continuous subspace of $H$} and
$\H_{\rm s}(H):=\H_{\rm p}(H)\oplus\H_{\rm sc}(H)$ for the \emph{singular subspace of $H$}. The sets
$\sigma_{\rm p}(H):=\sigma\big(H\vert_{\H_{\rm p}(H)}\big)$, 
$\sigma_{\rm ac}(H):=\sigma\big(H\vert_{\H_{\rm ac}(H)}\big)$,
$\sigma_{\rm sc}(H):=\sigma\big(H\vert_{\H_{\rm sc}(H)}\big)$, $\sigma_{\rm c}(H):=\sigma(H\vert_{\H_{\rm c}(H)})$
and $\sigma_{\rm s}(H):=\sigma(H\vert_{\H_{\rm s}(H)})$ are called \emph{pure point spectrum}, \emph{absolutely
continuous spectrum}, \emph{continuous spectrum}, and \emph{singular spectrum of $H$}, respectively. 

An important issue in spectral theory consists in determining the above spectral subspaces or subsets for
concrete selfadjoint operators. Under various assumptions this has been performed for important classes of
operators: Schr\"odinger and more general partial differential operators, Toeplitz operators, Wiener-Hopf
operators, and many others. Since the mathematical literature on this subject is considerable, it seems
pointless to try to indicate references.

In the present article we consider locally compact groups $X$, abelian or not, and convolution operators $H_\mu$, 
acting on $\ltwo(X)$, defined by suitable measures $\mu$ belonging to $\M(X)$, the Banach $^*$-algebra of complex
Radon measures on $X$. The case $\mu=\chi_S$, the characteristic function of a compact generating subset, leads
to Hecke operators associated to the left regular representation (notice that our groups need not to be discrete).
The precise definitions and statements are gathered in the next section. Essentially, our result consists in
determining subspaces $\K_\mu^1$ and $\K_\mu^2$ of $\ltwo(X)$, explicitly defined in terms of $\mu$ and the
family $\Hom(X,\R)$ of continuous group morphisms $\Phi:X\to\R$, such that $\H_{\rm p}(H_\mu)\subset\K_\mu^1$ and
$\H_{\rm s}(H_\mu)\subset\K_\mu^2$. The cases $\K_\mu^1=\{0\}$ or $\K_\mu^2=\{0\}$ are interesting; in the first
case $H_\mu$ has no eigenvalues, and in the second case $H_\mu$ is purely absolutely continuous. The subspaces
$\K_\mu^1$ and $\K_\mu^2$ can be calculated explicitly only in very convenient situations. Rather often we are
only able to show that they differ from $\H$.

In Section \ref{sec_main_result} we prove the results stated and discussed in Section \ref{sectionresult}. The
proofs rely on a modification of a positive commutator technique called \emph{the method of the weakly
conjugate operator}. This method, an unbounded version of the Kato-Putnam theorem \cite[Thm. XIII.28]{RS},
developed and used in various situations
\cite{BKM,Boutet/Mantoiu,Iftimovici/Mantoiu,Mantoiu/Pascu,Mantoiu/Richard,MRT,Ric06}, is recalled in Section
\ref{sec_weak}. 
The last section is devoted to examples.

We refer to \cite{BG00(1),BW05,BVZ,Bre1,Bre2,DS02,Georgescu/Golenia,GZ01,HRV93(1),HRV93(2),KSS06,Kes59,M/T,Si}
for some related works on the spectral theory of operators on groups and graphs. Some of these articles put into
evidence (Hecke-type) operators with large singular or singular continuous components. In \cite{MRT}, where
analogous technics are used, one gets restrictions on the singular spectrum for adjacency operators on certain
classes of graphs (which could be of non-Cayley type).

%-------------------------------------------------------------------------------------------------------------------
\section{The main result}\label{sectionresult}
\setcounter{equation}{0} 
%-------------------------------------------------------------------------------------------------------------------

We give in this section the statement of our main result for convolution operators on arbitrary locally compact
groups (LCG). The reader is referred to \cite{Fo95,HR63} for general information on the theory of LCG.

%-------------------------------------------------------------------------------------------------------------------
\subsection{Statement of the main result}
%-------------------------------------------------------------------------------------------------------------------

Let $X$ be a LCG with identity $e$, center $Z(X)$ and modular function $\Delta$. Let us fix a left Haar measure
$\lambda$ on $X$, using the notation $\d x:=\d\lambda(x)$. The associated right Haar measure $\rho$ is defined by
$\rho(E):=\lambda(E^{-1})$ for each Borel subset $E$ of $X$. Whenever $X$ is compact, $\lambda$ is normalized, \ie
$\lambda(X)=1$. On discrete groups the counting measure (assigning mass $1$ to every point) is considered. The
notation \emph{a.e.} stands for ``almost everywhere" and refers to the Haar measure $\lambda$. The Lebesgue space
$\lp(X)\equiv\lp(X,\d\lambda)$, $1\le p\le\infty$, of $X$ with respect to $\lambda$ is endowed with the usual norm
$$
\|f\|_p:=\(\int_X\d x\,|f(x)|^p\)^{1/p}.
$$

We are interested in convolution of functions by measures. Namely, we consider for every measure $\mu\in\M(X)$
and every function $f\in\lp(X)$, $1\le p<\infty$, the convolution of $\mu$ and $f$ given (essentially) by
$$
(\mu\ast f)(x):=\int_X\d\mu(y)\,f(y^{-1}x)\quad\hbox{for \emph{a.e.} }x\in X.
$$
It is known
\cite[Thm. 20.12]{HR63} that $\mu\ast f\in\lp(X)$ and that $\|\mu\ast f\|_p\le\|\mu\|\,\|f\|_p$, where
$\|\mu\|:=|\mu|(X)$ is the norm of the measure $\mu$. Since we are mainly concerned with the hilbertian theory,
we consider in the sequel the convolution operator $H_\mu$, $\mu\in\M(X)$, acting in the Hilbert space
$\H:=\ltwo(X)$:
$$
H_\mu f:=\mu\ast f,\quad f\in\H.
$$
The operator $H_\mu$ is bounded with norm $\|H_\mu\|\le\|\mu\|$, and it admits an adjoint operator $H_\mu^*$
equal to $H_{\mu^*}$, the convolution operator by $\mu^*\in\M(X)$ defined by $\mu^*(E)=\overline{\mu(E^{-1})}$.
If the measure $\mu$ is absolutely continuous w.r.t. the left Haar measure $\lambda$, so that
$\d\mu=a\,\d\lambda$ with $a\in\lone(X)$, then $\mu^*$ is also absolutely continuous w.r.t. $\lambda$ and
$\d\mu^*=a^*\d\lambda$, where $a^*(x):=\overline{a(x^{-1})}\Delta(x^{-1})$ for \emph{a.e.} $x\in X$. In such a
case we simply write $H_a$ for $H_{a\;\!\d\lambda}$. We shall always assume that $H_\mu$ is selfadjoint, \ie
that $\mu=\mu^*$.

Let $\U(\H)$ stands for the group of unitary operators in $\H$ and let $L:X\to\U(\H)$ be the left regular
representation of $X$. Then $H_\mu$ is equal to the strong operator integral
$$
H_\mu=\int_X\d\mu(y)\,L(y),
$$
and $\mu\mapsto H_\mu$ is the integrated form of $L$.

We recall that given two measures $\mu,\nu\in\M(X)$, their convolution $\mu\ast\nu\in \M(X)$ is defined by the
relation \cite[Eq. 2.34]{Fo95} generalizing the usual convolution of $\lone$-functions:
$$
\int_X\d(\mu\ast\nu)(x)\,g(x):=\int_X\int_X\d\mu(x)\d\nu(y)\,g(xy)\qquad\forall g\in C_0(X),
$$
where $C_0(X)$ denotes the $C^*$-algebra of continuous complex functions on $X$ decaying at infinity. The
inequality $\|\mu\ast\nu\|\le\|\mu\|\,\|\nu\|$ holds. 

Given $\mu\in\M(X)$, let $\varphi:X\to\R$ be such that the linear functional
$$
F:C_0(X)\to\C,\quad g\mapsto\int_X\d\mu(x)\,\varphi(x)g(x)
$$
is bounded. Then there exists a unique measure in $M(X)$ associated to $F$, due to the Riesz-Markov
representation theorem. We write $\varphi\mu$ for this measure, and we simply say that $\varphi$ is such that
$\varphi\mu\in\M(X)$. 

Let us call \emph{real character} any continuous group morphism $\Phi:X\to\R$. Their set forms a real vector
space $\Hom(X,\R)$, which can be infinite dimensional.

\begin{Definition}\label{admis}
Let $\mu=\mu^*\in\M(X)$. 
\begin{enumerate}
\item[(a)] A real character $\Phi$ is \emph{semi-adapted to $\,\mu$} if $\Phi\mu,\Phi^2\mu\in\M(X)$, and
$(\Phi\mu)\ast\mu=\mu\ast(\Phi\mu)$. The set of real characters that are semi-adapted to $\mu$ is denoted by
$\Hom^1_\mu(X,\R)$.
\item[(b)] A real character $\Phi$ is \emph{adapted to $\mu$} if $\Phi$ is semi-adapted to
$\mu,\Phi^3\mu\in\M(X)$, and $(\Phi\mu)\ast(\Phi^2\mu)=(\Phi^2\mu)\ast(\Phi\mu)$. The set of real characters
that are adapted to $\mu$ is denoted by $\Hom^2_\mu(X,\R)$.
\end{enumerate}
\end{Definition}

Let $\K^j_\mu:=\bigcap_{\Phi\in\Hom^j_\mu(X,\R)}\ker(H_{\Phi \mu})$, for $j=1,2$; then our main result is the
following.

\begin{Theorem}\label{princ}
Let $X$ be a LCG and let $\mu=\mu^*\in\M(X)$. Then
$$
\H_{\rm p}(H_\mu)\subset\K^1_\mu\qquad{\rm and}\qquad\H_{\rm s}(H_\mu)\subset\K^2_\mu.
$$
\end{Theorem}

A more precise result is obtained in a particular situation.

\begin{Corollary}\label{precis}
Let $X$ be a LCG and let $\mu=\mu^*\in\M(X)$. Assume that there exists a real character $\Phi$ adapted to $\mu$
such that $\Phi^2$ is equal to an nonzero constant on $\supp(\mu)$. Then $H_\mu$ has a purely absolutely
continuous spectrum, with the possible exception of an eigenvalue located at the origin, with eigenspace $\ker(H_\mu)=\ker(H_{\Phi\mu})$.
\end{Corollary}
Corollary \ref{precis} specially applies to adjacency operators on certain classes of Cayley graphs, which are
Hecke-type operators in the regular representation, thus convolution operators on discrete groups.

\begin{Remark}
Using the method of the weakly conjugate operator, some extra results (as a Limiting Absorption Principle, global
smooth operators, perturbations of $H_\mu$) can also be obtained. For simplicity we do not include them here, even
if they can be inferred quite straightforwardly from \cite{BKM} and \cite{Boutet/Mantoiu}. Improvements in the
assumptions are also possible, but with the cost of more complicated statements and proofs. Proposition 2.1 in
\cite{Boutet/Mantoiu} shows the generality of the method.
\end{Remark}

%-------------------------------------------------------------------------------------------------------------------
\subsection{Comments and remarks}\label{comarks}
%-------------------------------------------------------------------------------------------------------------------

(A) One obstacle in applying Theorem \ref{princ} is the fact that certain locally compact groups admit few nonzero
real characters, maybe none.

We say that $x\in X$ is \emph{compact}, and we write $x\in B(X)$, if the closed subgroup generated by $x$ is
compact. If the order of $x\in B(X)$ is finite, then $x$ is clearly a compact element (but in non-discrete groups
there could be others). Although $B(X)$ is the union of all the compact subgroups of $X$, it is in general neither
a subgroup, nor a closed set in $X$. We write $\B(X)$ for the closed subgroup generated by $B(X)$.

A continuous group morphism sends compact subgroups to compact subgroups. But the unique compact subgroup of $\R$
is $\{0\}$. Thus a real character on $X$ annihilates $\B(X)$. It is not clear in general that the ``smallness'' of
the vector space $\Hom(X,\R)$ is related to a tendancy for convolution operators on $X$ to have a substantial
singular subspace, but for certain classes of groups this is indeed the case. For example, if $X$ is compact, then
$X=\B(X)$, $\Hom(X,\R)=\{0\}$ and all operators $H_\mu$, $\mu\in\M(X)$, are pure point (see (B) below).

\medskip

(B) It is not at all exceptional for a convolution operator to have eigenvalues. For example, if a type I
representation $U$ is contained in the left regular representation $L$, then any function $a\in\lone(X)$ which is
transformed by (the integrated form of) $U$ into a compact operator will lead to a convolution operator $H_a$
having eigenvalues.

To consider just an exteme case, let us assume that $X$ is a CCR group and that $L$ is completely reducible. Then
$H_a$ can be written as a direct sum of compact operators, thus it has pure point spectrum. These conditions are
fulfilled in the very particular case of compact groups. Actually, in this case, the irreducible representations
are all finite-dimensional, so even convolution operators by elements of $\M(X)$ are pure point. 

\medskip

(C) The occurrence in Theorem \ref{princ} of the subspaces $\K^j_\mu$ is not as mysterious as it could seem at
first sight. For example, if $\mu=\delta_e$, then $\Phi\mu=0$ for any $\Phi\in\Hom(X,\R)$, so that $\K^j_\mu=\H$.
Accordingly $H_\mu=1$, with spectrum composed of the single eigenvalue $1$ with corresponding eigenspace $\H$.

Another simple example is obtained by considering $X$ compact. On one hand the single real character is $\Phi=0$,
with associated subspaces $\K^j_\mu=\H$ for any $\mu=\mu^*\in\M(X)$. On the other hand we know from (B) that
$\H_{\rm p}(H_\mu)$ is also equal to $\H$. 

If the support of $\mu$ is contained in a subgroup $Y$ of $X$ with $0<\lambda(Y)<\infty$, then a direct
calculation shows that the associated characteristic function $\chi_Y$ is an eigenvector of $H_\mu$ with
eigenvalue $\mu(Y)$. Actually, since $(\Phi\mu)(Y)=0$ for any $\Phi\in\Hom(X,\R)$ with $\Phi\mu\in\M(X)$,
$\C\chi_Y$ is contained in $\ker(H_{\Phi\mu})$.

%-------------------------------------------------------------------------------------------------------------------
\section{Proof of the main result}\label{sec_main_result}
\setcounter{equation}{0} 
%-------------------------------------------------------------------------------------------------------------------

The proof of Theorem \ref{princ} relies on an abstract method, that we briefly recall in a simple form.

%-------------------------------------------------------------------------------------------------------------------
\subsection{The method of the weakly conjugate operator}\label{sec_weak}
%-------------------------------------------------------------------------------------------------------------------

The method of the weakly conjugate operator works for unbounded operators, but for our purposes it will be enough
to assume $H$ bounded. It also produces estimations on the boundary values of the resolvent and information on
wave operators, but we shall only concentrate on spectral results.

We start by introducing some notations. The symbol $\H$ stands for a Hilbert space with scalar product $\<\:\!\cdot\:\!,\:\!\cdot\:\!\>$ and norm $\|\cdot\|$. Given two Hilbert spaces $\H_1$ and $\H_2$, we denote by
$\B(\H_1,\H_2)$ the set of bounded operators from $\H_1$ to $\H_2$, and put $\B(\H):=\B(\H,\H)$.  We assume that
$\H$ is endowed with a strongly continuous unitary group $\{W_t\}_{t\in\R}$. Its selfadjoint generator is denoted
by $A$ and has domain $\D(A)$. In most of the applications $A$ is unbounded.

\begin{Definition}\label{defC1}
A bounded selfadjoint operator $H$ in $\H$ belongs to $C^1(A;\H)$ if one of the following equivalent condition is
satisfied:
\begin{enumerate} 
\item[(i)] the map $\R\ni t\mapsto W_{-t}HW_t\in\B(\H)$ is strongly differentiable,
\item[(ii)] the sesquilinear form
$$
\D(A)\times\D(A)\ni(f,g)\mapsto i\<Hf,Ag\>-i\<Af,Hg\>\in\C
$$
is continuous when $\D(A)$ is endowed with the topology of $\H$.
\end{enumerate}
\end{Definition}

We denote by $B$ the strong derivative in (i) calculated at $t=0$, or equivalently the bounded selfadjoint
operator associated with the extension of the form in (ii). The operator $B$ provides a rigorous meaning to the
commutator $i[H,A]$. We shall write $B>0$ if $B$ is positive and injective, namely if $\<f,Bf\>>0$ for all
$f\in\H\setminus\{0\}$.

\begin{Definition}
The operator $A$ is \emph{weakly conjugate to} the bounded selfadjoint operator $H$ if $H\in C^1(A;\H)$ and
$B\equiv i[H,A]>0$.
\end{Definition}

For $B>0$ let us consider the completion $\BB$ of $\H$ with respect to the norm $\|f\|_\BB:=\<f,Bf\>^{1/2}$. The
adjoint space $\BB^*$ of $\BB$ can be identified with the completion of $B\H$ with respect to the norm $\|g\|_{\BB^*}:=\<g,B^{-1}g\>^{1/2}$. One has then the continuous dense embeddings
$\BB^*\hookrightarrow\H\hookrightarrow\BB$, and $B$ extends to an isometric operator from $\BB$ to $\BB^*$. Due to
these embeddings it makes sense to assume that $\{W_t\}_{t\in \R}$ restricts to a $C_0$-group in $\BB^*$, or
equivalently that it extends to a $C_0$-group in $\BB$. Under this assumption (tacitly assumed in the sequel) we
keep the same notation for these $C_0$-groups. The domain of the generator of the $C_0$-group in $\BB$ (resp.
$\BB^*$) endowed with the graph norm is denoted by $\D(A,\BB)$ (resp. $\D(A,\BB^*)$). In analogy with Definition
\ref{defC1} the requirement $B\in C^1(A;\BB,\BB^*)$ means that the map $\R\ni t\mapsto W_{-t}BW_t\in\B(\BB,\BB^*)$
is strongly differentiable, or equivalently that the sesquilinear form
$$
\D(A,\BB)\times\D(A,\BB)\ni(f,g)\mapsto i\<f,BAg\>-i\<Af,Bg\>\in\C
$$
is continuous when $\D(A,\BB)$ is endowed with the topology of $\BB$. Here, $\<\:\!\cdot\:\!,\:\!\cdot\:\!\>$
denotes the duality between $\BB$ and $\BB^*$. 

\begin{Theorem}\label{thmabstract}
Assume that $A$ is weakly conjugate to $H$ and that $B\equiv i[H,A]$ belongs to $C^1(A;\BB,\BB^*)$. Then the
spectrum of $H$ is purely absolutely continuous.
\end{Theorem}

Note that the method should be conveniently adapted when absolute continuity is expected only in a subspace of
the Hilbert space. This is the case considered in the sequel.

%-------------------------------------------------------------------------------------------------------------------
\subsection{Proof of Theorem \ref{princ}}
%-------------------------------------------------------------------------------------------------------------------

In this section we construct suitable weakly conjugate operators in the framework of section \ref{sectionresult},
and we prove our main result. For that purpose, let us fix a real character $\Phi\in\Hom(X,\R)$ and a measure
$\mu=\mu^*\in\M(X)$. We shall keep writing $\Phi$ for the associated operator of multiplication in $\H$. In most
of the applications this operator is unbounded; its domain is equal to $\D(\Phi)\equiv\{f\in\H\mid\Phi f\in\H\}$.

One ingredient of our approach is the fact that multiplication by morphisms behaves like a derivation with respect
to the convolution product: for suitable functions or measures $f,g:X\to\C$, one has
$\Phi(f\ast g)=(\Phi f)\ast g+f\ast(\Phi g)$. Using this observation we show in the next lemma that the commutator
$i[H_\mu,\Phi]$ (constructed as in Definition \ref{defC1}) is related to the operator $H_{\Phi \mu}$. This
provides a partial explanation of our choice of the ``semi-adapted" and ``adapted" conditions.

\begin{Lemma}\label{lemnoun}
\begin{enumerate}
\item[(a)] If $\Phi$ is semi-adapted to $\mu$, then $H_\mu\in C^1(\Phi;\H)$, and
$i[H_\mu,\Phi]=-iH_{\Phi \mu}\in\B(\H)$. Similarly, $-iH_{\Phi \mu}\in C^1(\Phi,\H)$, and
$i[-iH_{\Phi \mu},\Phi]=-H_{\Phi^2 \mu}\in\B(\H)$. Moreover, the equality $[H_\mu,H_{\Phi \mu}]=0$ holds. 
\item[(b)] If $\Phi$ is adapted to $\mu$, then $-H_{\Phi^2\mu}\in C^1(\Phi,\H)$, and the equality
$[H_{\Phi \mu},H_{\Phi^2 \mu}]=0$ holds.
\end{enumerate}
\end{Lemma}
 
\begin{proof}
(a) Let $\Phi$ be semi-adapted to $\mu$ and let $f\in\D(\Phi)$. Then one has $\mu,\Phi \mu\in\M(X)$ and
$f,\Phi f\in\H$. Thus $\mu\ast f\in\D(\Phi)$, and the equality $\Phi(H_\mu f)=(\Phi \mu)\ast f+\mu\ast(\Phi f)$
holds in $\H$. It follows that $i(H_\mu\Phi-\Phi H_\mu)$ is well-defined on $\D(\Phi)$ and is equal to
$-iH_{\Phi \mu}$. Hence Condition (ii) of Definition \ref{defC1} is fulfilled.

The proof that $-iH_{\Phi \mu}$ belongs to $C^1(\Phi,\H)$ and that $i[-iH_{\Phi \mu},\Phi]=-H_{\Phi^2 \mu}$ is
similar. Finally the equality $[H_\mu,H_{\Phi \mu}]=0$ is clearly equivalent to the requirement
$(\Phi \mu)\ast \mu=\mu\ast(\Phi \mu)$.

(b) The proof is completely analogous to that of point (a).
\end{proof}

If $\,\Phi$ is semi-adapted to $\mu$, we set $K:=i[H_\mu,\Phi]=-iH_{\Phi \mu}$ and
$L:=i[K,\Phi]=i[-iH_{\Phi \mu},\Phi]=-H_{\Phi^2 \mu}$ (for the sake of simplicity, we omit to write the dependence
of these operators in $\Phi$ and $\mu$). The first part of the previous lemma states that $H_\mu$ and $K$ belongs
to $C^1(\Phi;\H)$. In particular, it follows that $K$ leaves invariant the domain $\D(\Phi)$, and the operator 
$$
\A:=\hbox{$\frac12$}\(\Phi K+K\Phi\)
$$
is well-defined and symmetric on $\D(\Phi)$. Similarly, if $\Phi$ is adapted to $\mu$, the second part of the
lemma states that $L$ belongs to $C^1(\Phi;\H)$. Therefore the operator $L$ leaves $\D(\Phi)$ invariant, and the
operator
$$
\A':=\hbox{$\frac12$}\(\Phi L+L\Phi\)
$$
is well-defined and symmetric on $\D(\Phi)$. 

\begin{Lemma}\label{autoadjonction}
\begin{enumerate}
\item[(a)] If $\,\Phi$ is semi-adapted to $\mu$, then the operator $\A$ is essentially selfadjoint on $\D(\Phi)$.
The domain of its closure $A$ is $\D(A)=\D(\Phi K)=\{f\in\H\mid\Phi Kf\in\H\}$ and $A$ acts on $\D(A)$ as the
operator $\Phi K-\frac i2L$.
\item[(b)] If $\,\Phi$ is adapted to $\mu$, then the operator $\A'$ is essentially selfadjoint on $\D(\Phi)$. The
domain of its closure $A'$ is $\D(A')=\D(\Phi L)=\{f\in\H \mid \Phi Lf\in\H\}$.
\end{enumerate}
\end{Lemma}

\begin{proof}
One just has to reproduce the proof of \cite[Lemma 3.1]{Georgescu/Golenia}, replacing their couple $(N,S)$ by
$(\Phi,K)$ for the point (a) and by $(\Phi,L)$ for the point (b).
\end{proof}

In the next lemma we collect some results on commutators with $A$ or $A'$. The commutation relations
exhibited in Lemma \ref{lemnoun}, \ie $[H_\mu,K]=0$ if $\Phi$ is semi-adapted to $\mu$ and $[K,L]=0$ if $\Phi$ is
adapted to $\mu$, are essential.

\begin{Lemma}\label{Putnam}
If $\Phi$ is semi-adapted to $\mu$, then 
\begin{enumerate}
\item[(a)] The quadratic form $\D(A)\ni f\mapsto i\<H_\mu f,Af\>-i\<Af,H_\mu f\>$ extends uniquely to the bounded
form defined by the operator $K^2$,
\item[(b)] The quadratic form $\D(A)\ni f\mapsto i\<K^2f,Af\>-i\<Af,K^2f\>$ extends uniquely to the bounded
form defined by the operator $KLK+\hbox{$\frac12$}\(K^2L+LK^2\)$ (which reduces to $2KLK$ if $\Phi$ is
adapted to $\mu$),
\item[(c)] If $\Phi$ is adapted to $\mu$, then the quadratic form $\D(A')\ni f\mapsto i\<Kf,A'f\>-i\<A'f,Kf\>$
extends uniquely to the bounded form defined by the operator $L^2$.
\end{enumerate}
\end{Lemma}

The proof is straightforward. Computations may be performed on the core $\D(\Phi)$. These results imply that
$H_\mu\in C^1(A;\H)$, $K^2\in C^1(A;\H)$ and (when $\Phi$ is adapted) $K\in C^1(A';\H)$. Using these results we
now establish a relation between the kernels of the operators $H_\mu$, $K$ and $L$.

\begin{Lemma}\label{lemeq}
If $\Phi$ is semi-adapted to $\mu$, then one has
$$
\ker(H_\mu)\subset\H_{\rm p}(H_\mu)\subset\ker(K)\subset\H_{\rm p}(K).
$$
If $\Phi$ is adapted to $\mu$, one also has 
$$
\H_{\rm p}(K)\subset\ker(L)\subset\H_{\rm p}(L).
$$
\end{Lemma}

\begin{proof}
Let $f$ be an eigenvector of $H_\mu$. Due to the Virial Theorem \cite[Proposition 7.2.10]{ABG} and the fact that
$H_\mu$ belongs to $C^1(A;\H)$, one has $\<f,i[H_\mu,A]f\>=0$. It follows by Lemma \ref{Putnam}.(a) that
$0=\<f,K^2f\>=\|Kf\|^2$, \ie $f\in\ker(K)$. The inclusion $\H_{\rm p}(H_\mu)\subset \ker(K)$ follows. Similarly,
by using $A'$ instead of $A$ and Lemma \ref{Putnam}.(c) one gets (when $\Phi$ is adapted) the inclusion
$\H_{\rm p}(K)\subset\ker(L)$, and the lemma is proved.
\end{proof}

Assume now that $\Phi$ is semi-adapted to $\mu$. Then we can decompose the Hilbert space $\H$ into the direct sum
$\H=\K\oplus\G$, where $\K:=\ker(K)$ and $\G$ is the closure of the range $K\H$. It is easy to see that $H_\mu$
and $K$ are reduced by this decomposition and that their restrictions to the Hilbert space $\G$ are bounded
selfadjoint operators. In the next lemma we prove that this decomposition of $\H$ also reduces the operator $A$ if
$\Phi$ is adapted to $\mu$.

\begin{Lemma}\label{red}
If $\Phi$ is adapted to $\mu$, then the decomposition $\H=\K\oplus\G$ reduces the operator $A$. The restriction of
$A$ to $\G$ defines a selfadjoint operator denoted by $A_0$.
\end{Lemma}

\begin{proof}
We already know that on $\D(A)=\D(\Phi K)$ one has $A=\Phi K-\frac{i}{2}L$. By using Lemma \ref{lemeq} it follows
that $\K\subset\ker(A)\subset\D(A)$. Then one trivially checks that (i) $A\[\K\cap\D(A)\]\subset \K$, (ii)
$A\[\G\cap\D(A)\]\subset \G$ and (iii) $\D(A)=\[\K\cap\D(A)\]+\[\G\cap\D(A)\]$, which means that $A$ is reduced by
the decomposition $\H =\K\oplus\G$. Thus by \cite[Theorem 7.28]{Weidmann80} the restriction of $A$ to
$\D(A_0)\equiv\D(A)\cap\G$ is selfadjoint in $\G$. 
\end{proof}

\begin{proof}[Proof of Theorem \ref{princ}]
We know from Lemma \ref{lemeq} that $\H_{\rm p}(H_\mu)\subset\ker(-iH_{\Phi \mu})$ for each $\Phi\in\Hom^1_\mu(X,\R)$.
This obviously implies the first inclusion of the theorem.

Let us denote by $H_0$ and $K_0$ the restrictions to $\G$ of the operators $H_\mu$ and $K$. We shall prove in points
(i)-(iii) below that if $\Phi$ is adapted to $\mu$, then the method of the weakly conjugate operator, presented in
Section \ref{sec_weak}, applies to the operators $H_0$ and $A_0$ in the Hilbert space $\G$. It follows then that
$\G\subset\H_{\rm ac}(H_\mu)$, and \emph{a fortiori} that $\H_{\rm sc}(H_\mu)\subset\K=\ker(-iH_{\Phi \mu})$. Since
this result holds for each $\Phi\in\Hom^2_\mu(X,\R)$, the second inclusion of the theorem follows straightforwardly. 

(i) Lemma \ref{Putnam}.(a) implies that $i(H_0A_0-A_0H_0)$ is equal in the form sense to $K_0^2$ on
$\D(A_0)\equiv\D(A)\cap\G$. Therefore the corresponding quadratic form extends uniquely to the bounded form defined
by the operator $K_0^2$. This implies that $H_0$ belongs to $C^1(A_0;\G)$.

(ii) Since $B_0:=i[H_0,A_0]\equiv K_0^2>0$ in $\G$, the operator $A_0$ is weakly conjugate to $H_0$. So we define
the space $\BB$ as the completion of $\G$ with respect to the norm $\|f\|_\BB:=\<f,B_0f\>^{1/2}$. The adjoint space
of $\BB$ is denoted by $\BB^*$ and can be identified with the completion of $B_0\G$ with respect to the norm
$\|f\|_{\BB^*}:=\<f,B_0^{-1}f\>^{1/2}$. It can also be expressed as the closure of the subspace $K\H=K_0\G$ with
respect to the same norm $\|f\|_{\BB^*}=\big\||K_0|^{-1}f\big\|$. Due to Lemma \ref{Putnam}.(b) the quadratic form
$\D(A_0)\ni f\mapsto i\<B_0 fA_0 f\>-i\<A_0 f,B_0 f\>$ extends uniquely to the bounded form defined by the operator
$2K_0L_0K_0$, where $L_0$ is the restriction of $L$ to $\G$. We write $i[B_0,A_0]$ for this extension, which
clearly defines an element of $\B(\BB,\BB^*)$.

(iii) Let $\{W_t\}_{t\in\R}$ be the unitary group in $\G$ generated by $A_0$. We check now that this group extends
to a $C_0$-group in $\BB$. This easily reduces to proving that for any $t\in\R$ there exists a constant
$\textsc c(t)\ge0$ such that $\|W_tf\|_\BB\leq\textsc c(t)\|f\|_\BB$ for all $f\in\D(A_0)$. Due to point (ii) one
has for each $f\in\D(A_0)$
$$
\left\|W_t f\right\|^2_\BB=\<f,B_0f\>+\int_0^t\d\tau\<W_\tau f,i[B_0,A_0]W_\tau f\>
\leq\|f\|^2_\BB+2\|L_0\|\int_0^{|t|}\d\tau\left\|W_\tau f\right\|_\BB^2.
$$
Since $\G\hookrightarrow\BB$, the function $(0,|t|)\ni\tau\mapsto\|W_\tau f\|_\BB^2\in\R$ is bounded. Thus we get
the inequality $\|W_t f\|_\BB\leq\e^{|t|\|L_0\|}\|f\|_\BB$ by using a simple form of the Gronwall Lemma. Therefore
$\{W_t\}_{t\in\R}$ extends to a $C_0$-group in $\BB$, and by duality $\{W_t\}_{t\in\R}$ also defines a $C_0$-group
in $\BB^*$. This concludes the proof of the fact that $B_0$ extends to an element of $C^1(A_0;\BB,\BB^*)$. Thus
all hypotheses of Theorem \ref{thmabstract} are satisfied and this gives the result.
\end{proof}

\begin{proof}[Proof of Corollary \ref{precis}]
Since $L=-H_{\Phi^2\mu}$ is proportional with $H_\mu$, one has
$$
\ker(H_\mu)=\H_{\rm p}(H_\mu)\subset\H_{\rm s}(H_\mu)\subset\ker(H_{\Phi\mu})
$$
due to Lemma \ref{lemeq} and Theorem \ref{princ}. Using this with $\mu$ replaced by $i\Phi\mu$, one easily gets
the identity $\ker(H_{\Phi\mu})=\ker(H_\mu)$. Therefore
$$
\ker(H_\mu)=\H_{\rm p}(H_\mu)=\H_{\rm s}(H_\mu)=\ker(H_{\Phi\mu}),
$$
and the claim is proved.
\end{proof}

%-------------------------------------------------------------------------------------------------------------------
\section{Examples}\label{Examples}
\setcounter{equation}{0} 
%-------------------------------------------------------------------------------------------------------------------

%-------------------------------------------------------------------------------------------------------------------
\subsection{Perturbations of central measures}\label{central}
%-------------------------------------------------------------------------------------------------------------------

In this section, we exploit commutativity in a non-commutative setting by using central measures. The group $X$ is
assumed to be unimodular.

By definition, the \emph{central measures} are the elements of the center $Z[\M(X)]$ of the convolution Banach
$^*$-algebra $\M(X)$. They can be characterized by the condition $\mu(yEy^{-1})=\mu(E)$ for any $y\in X$ and any
Borel set $E\subset X$. The \emph{central} (or \emph{class}) \emph{functions} are the elements of
$Z[\M(X)]\cap\lone(X)=Z[\lone(X)]$. Thus a characteristic function $\chi_E$ is central iff $\lambda(E)<\infty$ and
$E$ is \emph{invariant} under all inner automorphisms. 

The relevant simple facts are the following: if $\mu$ is central, $\Phi\in\Hom(X,\R)$ and $\Phi\mu\in\M(X)$, then
$\Phi\mu$ is also central (this follows from the identity $\Phi(yxy^{-1})=\Phi(y)+\Phi(x)-\Phi(y)=\Phi(x)$,
$\forall x,y\in X$). On the other hand, if $\mu$ is arbitrary but supported on $\B(X)$, then $\Phi\mu=0$ for any
real character $\Phi$. Thus all the commutation relations in Definition \ref{admis} are satisfied, and one gets
from Theorem \ref{princ} the following result:

\begin{Corollary}\label{cabel}
Let $X$ be a unimodular LCG, let $\mu_0=\mu_0^*\in\M(X)$ be a central measure, and let $\mu_1=\mu_1^*\in\M(X)$
with $\supp(\mu_1)\subset\B(X)$. Then 
$$
\H_{\rm p}(H_{\mu_0+\mu_1})\subset\bigcap_{\substack{\Phi\in\Hom(X,\R)\\\Phi\mu_0,\Phi^2\mu_0\in\M(X)}}
\ker(H_{\Phi \mu_0})
$$
and
$$
\H_{\rm s}(H_{\mu_0+\mu_1})\subset\bigcap_{\substack{\Phi\in\Hom(X,\R)\\\Phi\mu_0,\Phi^2\mu_0,\Phi^3\mu_0\in\M(X)}}
\ker(H_{\Phi \mu_0}).
$$
\end{Corollary}

In order to get more explicit results, we restrict ourselves in the next section to a convenient class of LCG,
generalizing both abelian and compact groups.

%-------------------------------------------------------------------------------------------------------------------
\subsection{Convolution operators on central groups}\label{Central}
%-------------------------------------------------------------------------------------------------------------------

Following \cite{GM67}, we say that $X$ is \emph{central} (or \emph{of class $[Z]$}) if the quotient $X/Z(X)$ is
compact. Central groups possess a specific structure \cite[Thm. 4.4]{GM67}: If $X$ is central, then $X$ isomorphic
to a direct product $\R^d\times H$, where $H$ contains a compact open normal subgroup. 

\begin{Proposition}\label{centreaza}
Let $X$ be a central group and $\mu_0=\mu_0^*\in\M(X)$ a central measure such that $\supp(\mu_0)$ is compact and
not included in $\B(X)$. Let $\mu_1=\mu_1^*\in\M(X)$ with $\supp(\mu_1)\subset\B(X)$ and set $\mu:=\mu_0+\mu_1$.
Then $\H_{\rm ac}(H_\mu)\ne\{0\}$.
\end{Proposition}

\begin{proof}
Central groups are unimodular \cite[Prop. p. 366]{GM67(2)}, and $\Phi\mu,\Phi^2\mu,\Phi^3\mu\in\M(X)$ for any
$\Phi\in\Hom(X,\mathbb R)$, due to the hypotheses. Furthermore we know by \cite[Thm. 5.7]{GM67} that $B(X)=\B(X)$
is a closed normal subgroup of $X$ and that $X/\B(X)$ is isomorphic to the direct product $\R^d\times D$, where $D$
is a discrete torsion-free abelian group. But the groups $\R^d\times D$ are exactly those for which the real
characters separate points \cite[Cor. p. 335]{GM67}. Therefore for any $x\in\supp(\mu_0)\setminus\B(X)$ there
exists $\Phi\in\Hom(X,\R)$ such that $\Phi(x)\ne0$. Thus $H_{\Phi\mu}$ is a nonzero convolution operator, and the
claim follows by Corollary \ref{cabel}.
\end{proof}

In a central group $X$ there exists plenty of central compactly supported measures. For instance there always
exists in $X$ \cite[Thm. 4.2]{GM67} a neighbourhood base of $e$ composed of compact sets $S=S^{-1}$ which are
invariant (under the inner automorphisms), \ie central groups belong to the class [SIN]. Therefore the measures
$\mu_0=\chi_S\,\d\lambda$ satisfy $\mu_0=\mu_0^*$ and are subject to Proposition \ref{centreaza}. Actually this
also applies to central characteristic functions $\chi_S$ with ``large'' $S$, since in $X$ any compact set is
contained in a compact invariant neighbourhood of the identity \cite[Lemma p. 365]{GM67(2)}. One can also exihibit
central measures satisfying Proposition \ref{centreaza} defined by continuous functions. Indeed we know by
\cite[Thm. 1.3]{GM67(2)} that for any neighbourhood $U$ of the identity $e$ of a central group $X$ there exists a
non-negative continuous central function $a_U$, with $\supp(a_U)\subset U$ and $a_U(e)>0$.

A simple way to construct examples is as follows. Let $X:=K\times Y$, where $K$ is a compact group with Haar
measure $\lambda_K$ and $Y$ is an abelian LCG with Haar measure $\lambda_Y$. Clearly $X$ is central and
$\B(X)=K\times\B(Y)$. Let $\mathcal E$ be a finite family of invariant subsets of $K$ such that each
$E\in\mathcal E$ satisfies $\lambda_K(E)>0$ and $E^{-1}\in\mathcal E$. For each $E\in\mathcal E$, let $I_E$ be a
compact subset of $Y$ such that $\lambda_Y(I_E)>0$ and $(I_E)^{-1}=I_{E^{-1}}$. Suppose also that $I_{E_0}$ is
not a subset of $\B(Y)$ for some $E_0\in\mathcal E$. Then one easily shows that the set
$S:=\bigcup_{E\in\mathcal E}E\times I_E$ satisfies the following properties: $S$ is compact, $S=S^{-1}$, $S$ is
invariant, and $S$ not included in $\B(X)$. Thus $\H_{\rm ac}(H_{\chi_S+\mu_1})\ne\{0\}$ for any
$\mu_1=\mu_1^*\in\M(X)$ with $\supp(\mu_1)\subset K\times\B(Y)$, due to Proposition \ref{centreaza}.

The following two examples are applications of the preceding construction.

\begin{Example}
Let $X:=S_3\times\Z$, where $S_3$ is the symmetric group of degree $3$. The group $S_3$ has a presentation
$\<a,b\mid a^2,b^2,(ab)^3\>$, and its conjugacy classes are $E_1=E_1^{-1}=\{e\}$, $E_2=E_2^{-1}=\{a,b,aba\}$
and $E_3=E_3^{-1}=\{ab,ba\}$. Set $\mathcal E:=\{E_2,E_3\}$ and choose $I_{E_1},I_{E_2}$ two finite symetric
subsets of $\Z$, each of them containing at least two elements. Clearly these sets satisfy all the requirements
of the above construction. Thus $\H_{\rm ac}(H_{\chi_S})\ne\{0\}$ if $S:=\bigcup_{E\in\mathcal E}E\times I_E$.
\end{Example}

\begin{Example}
Let $X=\SU(2)\times\R$, where $\SU(2)$ is the group (with Haar measure $\lambda_2$) of $\,2\times2$ unitary
matrices of determinant $+1$. For each $\vartheta\in[0,\pi]$ let $C(\vartheta)$ be the conjugacy class of the
matrix $\diag(\e^{i\vartheta},\e^{-i\vartheta})$ in $\SU(2)$. A direct calculation (using for instance Euler
angles) shows that $\lambda_2\(\bigcup_{\vartheta\in J}C(\vartheta)\)>0$ for each $J\subset[0,\pi]$ with
nonzero Lebesgue measure. Set $E_1:=\bigcup_{\vartheta\in(0,1)}C(\vartheta)$,
$E_2:=\bigcup_{\vartheta\in(2,\pi)}C(\vartheta)$, $\mathcal E:=\{E_1,E_2\}$, $I_{E_1}:=(-1,1)$, and
$I_{E_2}:=(-3,-2)\cup(2,3)$. Clearly these sets and many others satisfy all the requirements of the above
construction. Thus $\H_{\rm ac}(H_{\chi_S})\ne\{0\}$ if $S:=\bigcup_{E\in\mathcal E}E\times I_E$.
\end{Example}

A nice example of a central group which is not the product of a compact and an abelian group can be found in
\cite[Ex. 4.7]{GM71}. 

In a simple situation one even gets purely absolutely continuous operators; this should be
compared with the discussion in Section \ref{comarks}:

\begin{Example}
Let $X$ be a central group, let $z\in Z(X)\setminus\B(X)$, and set $\mu:=\delta_z+\delta_{z^{-1}}+\mu_1$ for some $\mu_1=\mu_1^*\in\M(X)$ with $\supp(\mu_1)\subset\B(X)$. Then $\mu$ satisfies the hypotheses of Proposition
\ref{centreaza}, and we can choose $\Phi\in\Hom(X,\R)$ such that $\Phi(z)=\frac12\Phi(z^2)\ne0$ (note in particular
that $z\notin\B(X)$ iff $z^2\notin\B(X)$ and that $\Phi\mu_1=0$). Thus $\H_{\rm s}(H_\mu)\subset\ker(H_{\Phi\mu})$.
But $f\in\H$ belongs to $\ker(H_{\Phi\mu})=\ker\big(H_{\Phi(\delta_z+\delta_{z^{-1}})}\big)$ iff
$f(z^{-1}x)=f(zx)$ for {\em a.e.} $x\in X$. This periodicity w.r.t. the non-compact element $z^2$ easily implies
that the $\ltwo$-function $f$ should vanish {\em a.e.} and thus that $\H_{\rm ac}(H_\mu)=\H$.
\end{Example}

%-------------------------------------------------------------------------------------------------------------------
\subsection{Abelian groups}\label{abelian}
%-------------------------------------------------------------------------------------------------------------------

We consider in this section the case of \emph{locally compact abelian groups} (LCAG), whose theory can be found in
the monograph \cite{HR63}. LCAG are particular cases of central groups. Their convolution algebra $\M(X)$ is
abelian, so spectral results on convolution operators can be deduced from the preceding section. We shall not
repete them here, but rather invoke duality to obtain properties of a class of multiplication operators on the
dual group $\widehat X$.

Let $X$ stands for a LCAG with elements $x,y,z,\ldots$, and let $\widehat X$ be the dual group of $X$, \ie the set
of characters of $X$ endowed with the topology of compact convergence on $X$. The elements of $\widehat X$ are
denoted by $\xi,\eta,\zeta,\ldots$ and we shall use the notation $\<x,\xi\>$ for the expression $\xi(x)$.
The Fourier transform $m$ of a measure $\mu\in\M(X)$ is given by
$$
m(\xi)\equiv[\F(\mu)](\xi):=\int_X\d\mu(x)\,\overline{\<x,\xi\>},\quad\xi\in\widehat X.
$$
We recall from \cite[Thm. 23.10]{HR63} that $m$ belongs to the $C^*$-algebra $BC(\widehat X)$ of bounded
continuous complex functions on $\widehat X$, and that $\|m\|_\infty\le\|\mu\|$ (showing that the bound
$\|H_\mu\|\le\|\mu\|$ is not optimal in general). Actually the subspace $\F(\M(X))$ is dense in $BC(\widehat X)$,
and the subspace $\F(\lone(X))$ is densely contained in $C_0(\widehat X)$, the ideal of $BC(\widehat X)$ composed
of continuous complex functions on $\widehat X$ vanishing at infinity. For a suitably normalized Haar measure on
$\widehat X$, the Fourier transform also defines a unitary isomorphism from $\H$ onto $\ltwo(\widehat X)$, which we
denote by the same symbol. It maps unitarily $H_\mu$ on the operator $M_m$ of multiplication with $m=\F(\mu)$.
Moreover $\mu=\mu^*$, iff $m$ is real, and
$$
\sigma(H_\mu)=\sigma(M_m)=\overline{m(\widehat X)},\quad\sigma_{\rm p}(H_\mu)=\sigma_{\rm p}(M_m)
=\overline{\left\{s\in\R\mid\lambda_\land\(m^{-1}(s)\)>0\right\}},
$$
where $\lambda_\land$ is any Haar measure on $\widehat X$.

This does not solve the problem of determining the nature of the spectrum, at least for three reasons. First, simple
or natural conditions on $\mu$ could be obscured when using the Fourier transform; the function $m=\F(\mu)$ could be
difficult to compute or to evaluate. Second, the dual group $\widehat X$ can be complicated. We are not aware of
general results on the nature of the spectrum of multiplication operators on LCAG. Third, even for $\widehat X=\R^d$,
the spectral theory of multiplication operators is quite subtle. For the particular case $\widehat X=\R^d$, one
finds in \cite[Sec. 7.1.4 \& 7.6.2]{ABG} refined results both on the absolute continuity and on the occurence of
singular continuous spectrum for multiplication operators. 

Let us recall that there is an almost canonical identification of $\Hom(X,\R)$ with the vector space
$\Hom(\R,\widehat X)$ of all continuous one-parameter subgroups of $\widehat X$. For a given real character $\Phi$,
we denote by $\varphi\in\Hom(\R,\widehat X)$ the unique element satisfying
$$
\<x,\varphi(t)\>=\e^{it\Phi(x)},\qquad\forall t\in\R,~x\in X.
$$

\begin{Definition}
The function $m:\widehat X\to\C$ is \emph{differentiable at $\xi\in\widehat X$ along the one-parameter subgroup $\varphi\in\Hom(\R,\widehat X)$} if the function $\R\ni t\mapsto m(\xi+\varphi(t))\in\C$ is differentiable at $t=0$.
In such a case we write $\(d_\varphi m\)(\xi)$ for $\frac\d{\d t}\;\!m(\xi+\varphi(t))\big\vert_{t=0}$. Higher
order derivatives, when existing, are denoted by $d_{\varphi}^km$, $k\in\N$.
\end{Definition}

This definition triggers a formalism which has some of the properties of the differential calculus on $\R^d$.
However a differentiable function might not be continuous. Moreover, if $\widehat X$ is totally disconnected, then
the theory is trivial: Every complex function defined on $\widehat X$ is differentiable with respect to the single
trivial element of $\Hom(\R,\widehat X)$, and the derivative is always zero. 
If $\mu\in\M(X)$ is such that $\Phi\mu\in\M(X)$, then \cite[p. 68]{Ri53} $m=\F(\mu)$ is differentiable at any
point $\xi$ along the one-parameter subgroup $\varphi$ and $-i\F(\Phi\mu)=d_\varphi m$.

Let us fix a bounded continuous function $m:\widehat X\to\R$ such that $\F^{-1}(m)\in\M(X)$. We say that the
one-parameter subgroup $\varphi:\R\to\widehat X$ is in $\Hom_m^1(\R,\widehat X)$ if $m$ is twice differentiable
w.r.t. $\varphi$ and $d_\varphi m,d^2_\varphi m\in\F(\M(X))$. If, in addition, $m$ is thrice differentiable w.r.t.
$\varphi$ and $d^3_\varphi m\in\F(\M(X))$ too, we say that $\varphi$ belongs to $\Hom_m^2(\R,\widehat X)$. Then
next result follows directly from Corollary \ref{cabel}.

\begin{Corollary}\label{babel}
Let $X$ be a LCAG and let $m_0,m_1$ be real functions with $\F^{-1}(m_0),\F^{-1}(m_1)\in\M(X)$ and
$\supp(\F^{-1}(m_1))\subset\B(X)$. Then 
$$
\H_{\rm p}(M_{m_0+m_1})\subset\bigcap_{\varphi\in\Hom_{m_0}^1(\R,\widehat X)}\ker(M_{d_\varphi m_0})
$$
and
$$
\H_{\rm s}(M_{m_0+m_1})\subset\bigcap_{\varphi\in\Hom^2_{m_0}(\R,\widehat X)}\ker(M_{d_\varphi m_0}).
$$
\end{Corollary}

It is worth noting that for $X$ abelian, the following assertions are equivalent
\cite[Thm. 24.34 \& Cor. 24.35]{HR63}: (i) $\B(X)=\{e\}$, (ii) the real characters separate points, (iii) the dual
group $\widehat X$ is connected, (iv) $X$ is isomorphic to $\R^d\times D$, where $D$ is a discrete torsion-free
abelian group. 

Up to our knowledge, Corollary \ref{babel} is not known in the present generality. It is a by-product of a theory
working in a non-commutative framework and it is obviously far from being optimal. We hope to treat the spectral
analysis of (unbounded) multiplication operators on LCAG in greater detail in a forthcoming publication.

One may interpret our use of $\Hom(X,\R)$ in Theorem \ref{princ} as an attempt to involve ``smoothness'' and
``derivatives'' in spectral theory for groups which might not be abelian or might not have a given Lie structure.

%-------------------------------------------------------------------------------------------------------------------
\subsection{Semidirect products}\label{semi}
%-------------------------------------------------------------------------------------------------------------------

Let $N,G$ be two discrete groups with $G$ abelian (for which we use additive notations), and let $\tau:G\to\Aut(N)$
be a group morphism. Let $X:=N\times_\tau G$ be the $\tau$-semidirect produt of $N$ by $G$. The multiplication in
$X$ is defined by
$$
(n,g)(m,h):=(n\tau_g(m),g+h),
$$
so that
$$
(n,g)^{-1}=(\tau_{-g}(n^{-1}),-g).
$$
In the sequel we only consider real characters $\Phi\in\Hom(X,\R)$ of the form $\Phi=\phi\circ\pi$, where
$\phi\in\Hom(G,\R)$ and $\pi:X\to G$ is the canonical morphism given by $\pi(n,g):=g$.

\begin{Proposition}\label{bouboulina}
Let $a_0=a_0^*:X\to\C$ have a finite support and satisfy
\begin{equation}\label{bequilles}
\sum_{\substack{n_1,n_2\in N\\m=n_1\tau_{g_1}(n_2)}}a_0(n_1,g_1)a_0(n_2,g_2)
=\sum_{\substack{n_1,n_2\in N\\m=n_1\tau_{g_2}(n_2)}}a_0(n_1,g_2)a_0(n_2,g_1),
\quad\forall g_1,g_2\in G,~\forall m\in N.
\end{equation}
Let $a_1=a_1^*:X\to\C$ have a finite support contained in $\B(X)$ and be such that $a_1\ast a_0=a_0\ast a_1$.
Then
$$
\H_{\rm s}(H_{a_0+a_1})\subset\bigcap_{\phi\in\Hom(G,\R)}\ker\big(H_{(\phi\circ\pi)a_0}\big).
$$
\end{Proposition}

\begin{proof}
Since $a:=a_0+a_1$ has a finite support, we only have to check that
\begin{equation}\label{asta}
\Phi a\ast a-a\ast\Phi a=\Phi a\ast\Phi^2a-\Phi^2a\ast\Phi a=0
\quad\textrm{for any }\Phi=\phi\circ\pi,~\phi\in\Hom(G,\R).
\end{equation}
Since $a_1\ast a_0=a_0\ast a_1$ and $\Phi a_1=0$ for each $\Phi\in\Hom(X,\R)$, we are easily reduced to check
\eqref{asta} only for $a_0$. Let $(m,h)\in X$; then a direct calculation gives
$$
(\Phi a_0\ast a_0-a_0\ast\Phi a_0)(m,h)=\sum_{g\in G}\phi(2g-h)\sum_{\substack{n_1,n_2\in N\\m=n_1\tau_g(n_2)}}
a_0(n_1,g)a_0(n_2,h-g).
$$
This leads to the identity
$$
(\Phi a_0\ast a_0-a_0\ast\Phi a_0)(m,h)=-(\Phi a_0\ast a_0-a_0\ast\Phi a_0)(m,h)
$$
by using Condition \eqref{bequilles} and the change of variable $g':=h-g$. Thus
$\Phi a_0\ast a_0-a_0\ast\Phi a_0=0$.

By a similar argument one obtains that $\Phi a_0\ast \Phi^2a_0-\Phi^2a_0\ast\Phi a_0=0$ (the extra factor
involved in this computation is symmetric with respect to the change of variables).
\end{proof}

The proposition tells us that $H_a$ has a non-trivial absolutely continuous component if there exists a real
character $\phi\in\Hom(G,\R)$ such that $(\phi\circ\pi)a\ne0$. Consequently, as soon as $\supp(a)$ is not
included in $N\times\B(G)$, we are done. For instance if $G=\Z^d$, we simply have to ask for the existence
of an element $(n,g)\in\supp(a)$ with $g\ne0$. In the remaining part we indicate several situations to which
Proposition \ref{bouboulina} applies; the perturbation $a_1$ is left apart for simplicity.

\begin{Procedure}\label{garp_garp}
Let $G_0$ be a finite subset of $G$ such that $G_0=-G_0$. For each $g\in G_0$ let $N_g$ be a finite subset of
$N$ such that $\tau_g\big(N_{-g}\big)=N_g^{-1}$ for each $g\in G_0$. Set
$$
S:=\bigsqcup_{g\in G_0}N_g\times\{g\}.
$$
This is a convenient description of the most general finite subset of $X$ satisfying $S^{-1}=S$ (which is
equivalent to $\chi_S=\chi_S^*$). Condition \eqref{bequilles} amounts to
\begin{equation}\label{quation}
\#\{(n_1,n_2)\in N_{g_1}\times N_{g_2}\mid m=n_1\tau_{g_1}(n_2)\}
=\#\{(n_1,n_2)\in N_{g_2}\times N_{g_1}\mid m=n_1\tau_{g_2}(n_2))\}
\end{equation}
for each $g_1,g_2\in G_0$ and $m\in N$. Under these assumptions Proposition \ref{bouboulina} applies and
$H_{\chi_S}$ has a non-trivial absolutely continuous component if $G_0\cap[G\setminus\B(G)]\ne\varnothing$.
\end{Procedure}

One can ensure in various situations that $S$ is a system of generators (which is needed to assign a
\emph{connected} Cayley graph to $(X,S)$). This happens, for instance, if $G_0$ generates $G$,
$\cup_{g\in G_0}N_g$ generates $N$ and the unit $e$ of $N$ belongs to $N_g$ for each $g\in G_0$.

\begin{Example}
Let $N:=S_3=\<a,b\mid a^2,b^2,(ab)^3\>$, $G:=\Z$, $G_0:=\{-1,1\}$, $N_{-1}:=\{a,aba\}$, $N_{1}:=\{a,b\}$,
and $\tau_g(n):=a^gna^{-g}$ for each $g\in\Z$ and $n\in S_3$. Then direct calculations show that all the
assumptions in Procedure \ref{garp_garp} are verified. Hence $\H_{\rm ac}(H_{\chi_S})\ne\{0\}$ for
$X:=S_3\times_\tau\Z$ if $S=\{(a,-1),(aba,-1),(a,1),(b,1)\}$. Actually, by applying Corollary \ref{precis}
with $\Phi(n,g):=g$, one finds that the single possible component of the singular spectrum of $H_{\chi_S}$
is an eigenvalue located at the origin. However a careful inspection shows us that $H_{\chi_S}$ is
injective, and thus that $\H_{\rm ac}(H_{\chi_S})=\H$.
\end{Example}

A rather simple (but not trivial) possibility consists in taking $N_g\equiv N_0$ independent on $g\in G_0$
in Procedure \ref{garp_garp}. In such a case $S=N_0\times G_0$ and $\chi_S=\chi_{N_0}\otimes\chi_{G_0}$
(which does not implies in general that $H_{\chi_S}$ is a tensor product of operators). If we also assume
that $N_0$ is invariant under the set $\{\tau_g\mid g\in G_0\}$, then all the necessary assumptions are
satisfied. For instance, the invariance of $N_0$ permits to define a bijection between the two sets in
Formula \eqref{quation}. A particular case would be to choose $G:=\Z^d$, with
$G_0:=\{(\pm1,\ldots,0),\ldots,(0,\ldots,\pm1)\}$. By using the real morphism $\phi:\Z^d\to\R$ defined by
$\phi(\pm1,\ldots,0)=:\pm1,\ldots,\phi(0,\ldots,\pm1)=:\pm1$, one can apply Corollary \ref{precis} to
conclude that, except the possible eigenvalue $0$, $H_{\chi_S}$ is purely absolutely continuous.

The following two examples are applications of the preceding construction.

\begin{Example}
We consider a simple type of wreath product. Take $G$ a discrete abelian group and put $N:=R^J$, where $R$ is an
arbitrary discrete group and $J$ is a finite set on which $G$ acts by $(g,j)\mapsto g(j)$. Then
$\tau_g\big(\{r_j\}_{j\in J}\big):=\{r_{g(j)}\}_{j\in J}$ defines an action of $G$ on $R^J$, thus we can
construct the semidirect product $R^J\times_\tau G$. If $G_0=-G_0\subset G$ and $R_0=R_0^{-1}\subset R$ are
finite subsets with $G_0\cap[G\setminus\B(G)]\ne\varnothing$, then $N_0:=R_0^J$ satisfies all the required
conditions. Thus $\H_{\rm ac}(H_{\chi_S})\ne\{0\}$ if $S:=N_0\times G_0$.
\end{Example}

\begin{Example}
Let $G$ be a discrete abelian group, let $N$ be the free group generated by the family $\{a_1,\ldots,a_n\}$, and
set $N_0:=\big\{a^{\pm1}_1,\ldots,a_k^{\pm 1}\big\}$. Choose a finite set $G_0=-G_0\subset G$ with $G_0\cap[G\setminus\B(G)]\ne\varnothing$ and an action $\tau$ on $N$ such that the conditions on
$S:=N_0\times G_0$ are satisfied (for instance $\tau_g$, $g\in G_0$, may act by permutation on the generators).
Then $H_{\chi_S}$ has a non-trivial absolutely continuous part.
\end{Example}

Virtually the methods of this article could also be applied to non-split group extensions.

%-------------------------------------------------------------------------------------------------------------------
\section*{Acknowledgements}
%-------------------------------------------------------------------------------------------------------------------

M.M. acknowledges support from the contract 2-CEx 06-11-34. R.T.d.A. thanks the Swiss National Science Foundation
for financial support. Part of this work was done while M.M. was visiting the University of Paris XI. He would
like to thank Professor B. Helffer for his kind hospitality. Both authors are grateful to V. Georgescu and S.
Richard for useful discussions.

%-------------------------------------------------------------------------------------------------------------------
%Bibliography
%-------------------------------------------------------------------------------------------------------------------

\end{document}